\documentclass{ws-procs975x65}

\begin{document}

\title{$n$-DBI GRAVITY IN A NUTSHELL}

\author{F. S. COELHO$^*$ and C. HERDEIRO}

\address{Departamento de F\'isica da Universidade de Aveiro and I3N,\\
Campus de Santiago, 3810-183 Aveiro, Portugal\\
$^*$E-mail: flavio@physics.org}

\author{S. HIRANO and Y. SATO}

\address{Department of Physics, Nagoya University,\\
Nagoya 464-8602, Japan}

\begin{abstract}
We present a new model of gravity which explicitly breaks Lorentz-invariance by the introduction of a unit time-like vector field, thereby giving rise to an extra (scalar) degree of freedom. We discuss its cosmology, exact solutions and the dynamics of the scalar mode. We show that it predicts inflation without an inflaton and admits the black hole solutions of General Relativity (GR). We argue that the scalar mode is well behaved and contains none of the pathologies previously found in similar models.
\end{abstract}


\bodymatter

\section{$n$-DBI Gravity}
The action for $n$-DBI \cite{Herdeiro:2011im} without matter is
\begin{equation}
S=-\frac{3\lambda}{4\pi G_N^2}\int d^4x\sqrt{-g}\left\{\sqrt{1+\frac{G_N}{6\lambda}\left(R+\mathcal{K}\right)}-q\right\}\,,\label{action}
\end{equation}
where $\lambda$ and $q$ are arbitrary constants, $G_N$ is Newton's constant, $R$ is the Ricci scalar and $\mathcal{K}$ is defined through a unit, everywhere time-like, vector field $n^\mu$ as $\mathcal{K}=-2\nabla_\mu(n^\mu\nabla_\nu n^\nu)$. Concretely, we work in the Arnowitt-Deser-Misner (ADM) formalism and foliate space-time by space-like hypersurfaces orthogonal to $n^\mu$.

\medskip

The introduction of the vector field allows the field equations to remain second order in time derivatives, while becoming higher order in spatial derivatives. Furthermore, it breaks the general diffeomorphisms group of General Relativity down to the subgroup of {\it foliation-preserving diffeomorphisms}:
\begin{equation}
t\rightarrow t+\xi^0(t)\,,\qquad x^i\rightarrow x^i+\xi^i(t,x)\,.
\end{equation}
This gives rise to a scalar degree of freedom in addition to the two tensor modes of General Relativity, very much alike Ho\v{r}ava-Lifshitz (HL) gravity.

\medskip

General Relativity is recovered by taking the double limit $\lambda\rightarrow\infty, q\rightarrow1$ while keeping the product $\lambda(q-1)$ fixed, which yields Einstein's equations with a cosmological constant $\Lambda_C=6\lambda(q-1)/G_N^2$. Indeed, for weak curvature, the leading order term in Eq.~\ref{action} is the Einstein-Hilbert action with the Gibbons-Hawking-York boundary term.

\section{Cosmology}
The original motivation for this model \cite{Herdeiro:2011km} comes from the realization that for a conformally flat Universe it yields a Dirac-Born-Infeld type conformal scalar theory. Concretely, for the Friedmann-Robertson-Walker ansatz
\begin{equation}
ds^2=l_P^2\phi^2\left(-d\tau^2+\delta_{ij}dx^idx^j\right)\,,
\end{equation}
where $l_P$ is the Planck length, the conformal factor $\phi$ is governed by
\begin{equation}
\frac{1}{2}\dot{\phi}^2+V(\phi)=0\,,\qquad V(\phi)=-\frac{1}{2}\lambda\phi^4\left[1-\left(q+\frac{\epsilon}{\lambda\phi^4}\right)^{-2}\right]\,.
\end{equation}
The integration constant $\epsilon$ can be shown to be the energy density of radiation, so we are actually dealing with a Universe permeated by a perfect fluid obeying $\rho=3p$.

\begin{figure}[h]
\begin{center}
\parbox{2.3in}{\epsfig{figure=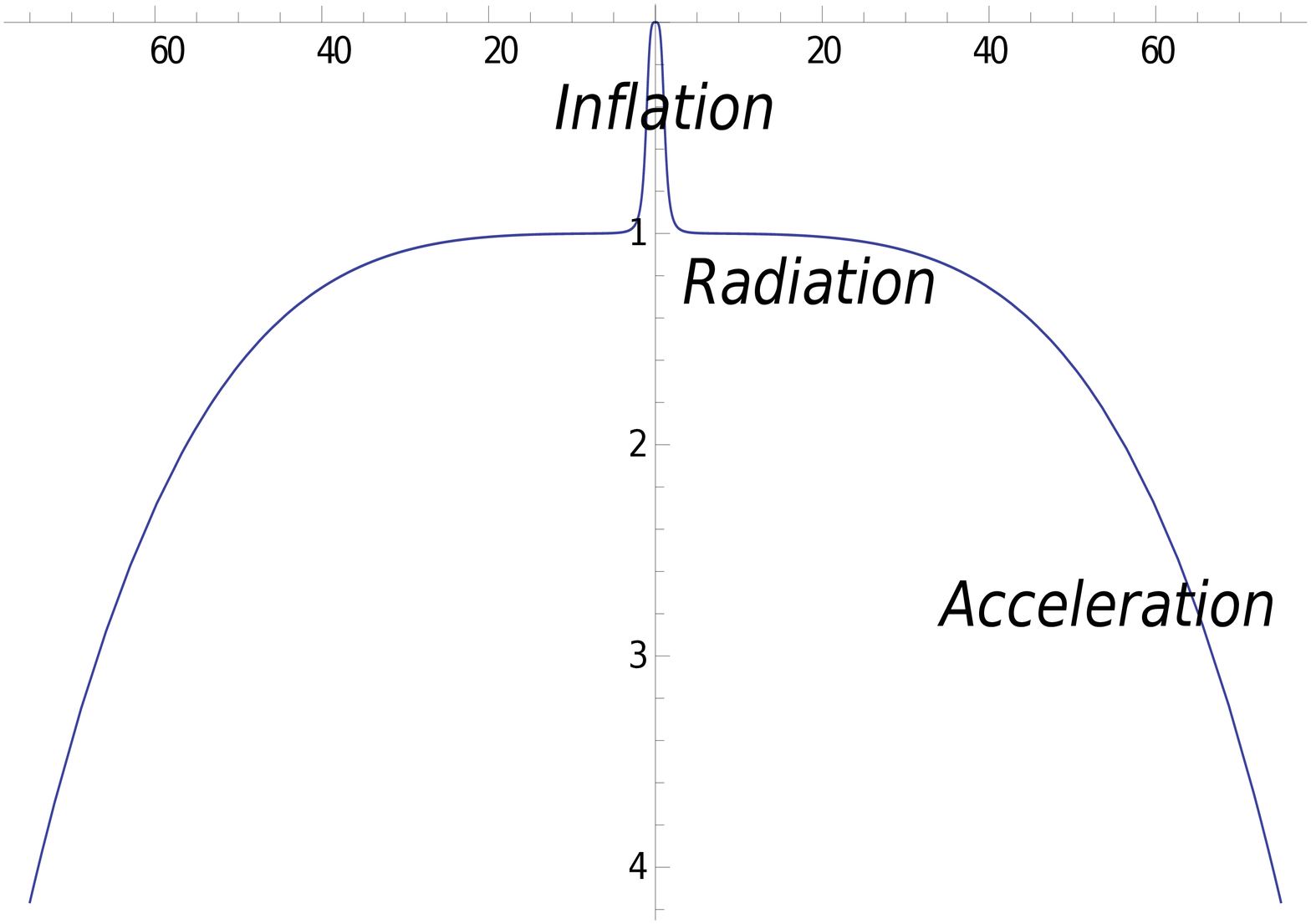,width=2.3in}
}
\hspace*{20pt}
\parbox{2.3in}{\epsfig{figure=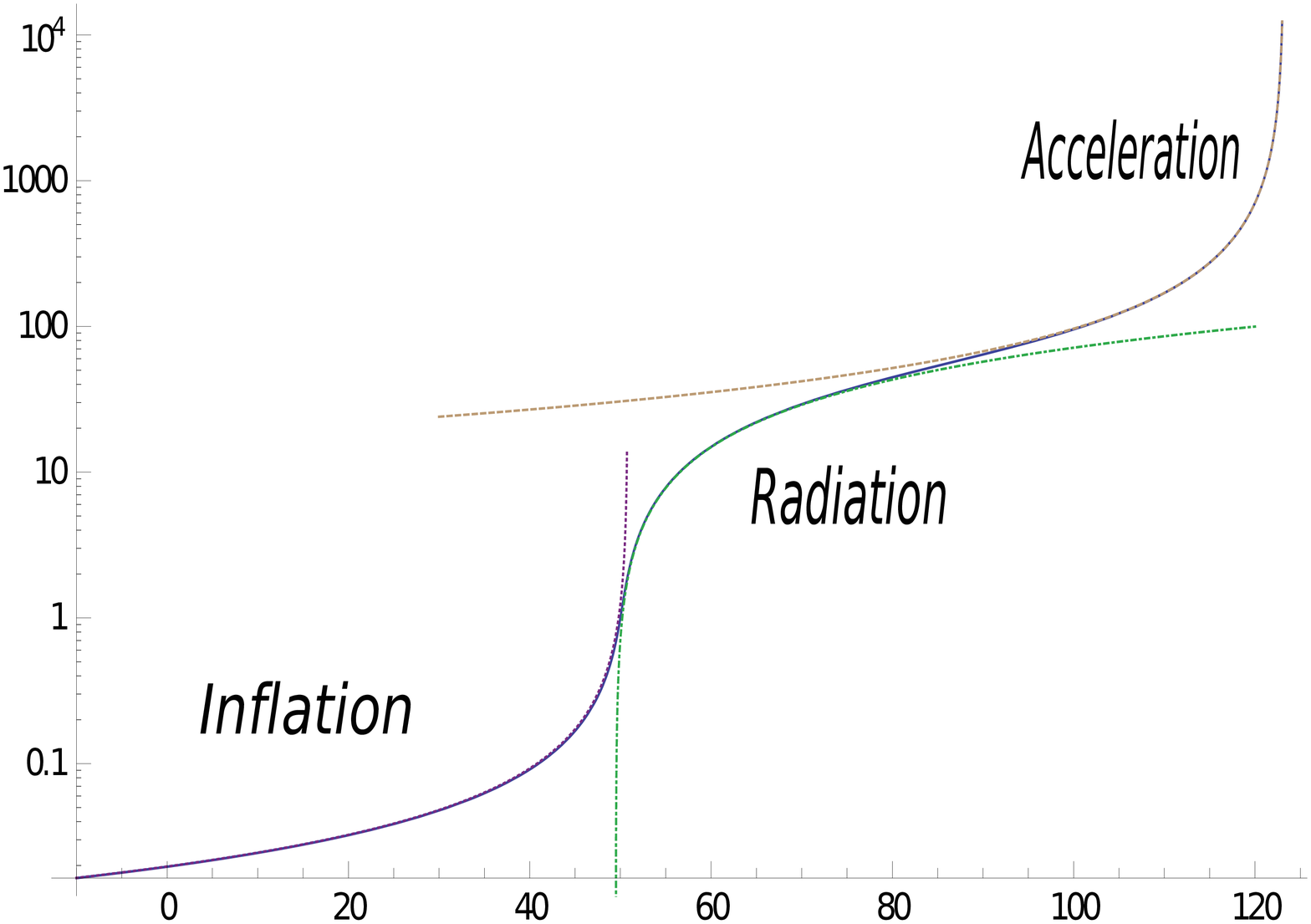,width=2.3in}
}
\caption{{\it Left:} The effective potential $V(\phi)$. {\it Right:} Time evolution of the conformal factor $\phi(\tau)$.}
\label{fig1}
\end{center}
\end{figure}

As we can see from Fig.~\ref{fig1}, this model naturally results in primordial inflation, followed by a radiation dominated epoch and subsequent accelerated expansion at late times. Matter can also be included in this picture without significant qualitative changes. The current value of the cosmological constant, approximately $10^{-12}$ GeV, and an energy scale of inflation of the order $10^{15}$ GeV constrain the free parameters $\lambda\sim10^{-8}$, $q-1\sim 10^{-110}$. The required fine-tuning is of the same order as in standard cosmology, and generates a large hierarchy between the two effective cosmological constants, $1/\sqrt{1-q^{-2}}$. It is possible that the scalar mode is in fact acting as an inflaton. We anticipate that it could source scalar perturbations in the Cosmic Microwave Background, and hope to report on that soon.

\section{Black Hole Solutions}
The field equations of $n$-DBI simplify tremendously if we restrict to solutions with constant $R+\mathcal{K}$, in which case one can prove the following theorem and corollary \cite{Herdeiro:2011im}:
\begin{theorem}Any solution of EinsteinÕs gravity with cosmological constant plus matter, admitting a foliation with constant $R+\mathcal{K}$, is a solution of $n$-DBI gravity.
\end{theorem}
\begin{corollary}\label{cor}
Any Einstein space admitting a foliation with constant $^{(3)}R-N^{-1}\Delta N$ (where $N$ is the lapse and $^{(3)}R$ the Ricci scalar of the 3-dimensional hypersurfaces), is a solution of n-DBI gravity.
\end{corollary}

By requiring spherical symmetry, one can explicitly obtain the Schwarzschild, Reissner-Nordstr\"om and (anti) de Sitter black hole solutions, albeit in an unusual set of coordinates. Unlike General Relativity, however, the cosmological constant is not determined at the level of the action but appears instead as an integration constant. Corollary \ref{cor} can be interpreted as the {\it maximal slicing} gauge condition common in Numerical Relativity, and it is then straightforward to show that the Kerr metric in Boyer-Lindquist coordinates is also a solution of $n$-DBI gravity \cite{Coelho:2012kerr}.

\section{The Scalar Mode}
The existence of an extra scalar degree of freedom was established through Dirac's theory of constrained systems \cite{Coelho:2012sg}. Whereas in GR the Hamiltonian constraint is automatically preserved, here its time evolution gives an equation for the scalar mode. By studying perturbations around flat space, we concluded that this mode is {\it non propagating} and carries no energy. We further argued that it is free from the pathologies that afflict some versions of HL gravity, namely the issues of vanishing lapse \cite{Henneaux:2009zb}, short-distance instability and strong coupling \cite{Blas:2009yd}. This good behaviour stems from the non-linear lapse terms in the action, also introduced in healthy extensions of HL gravity \cite{Blas:2009qj}. However, its true nature and possible observational consequences remain to be understood, and will be the subject of future work.

\section*{Acknowledgments}
F.C. is funded by FCT through the grants SFRH/BD/60272/2009. This work was partially supported by the Grant-in-Aid for Nagoya University Global COE Program (G07), by FCT (Portugal) through the project PTDC/FIS/116625/2010 and by the Marie Curie Action NRHEPÐ295189-FP7-PEOPLE-2011-IRSES.

\bibliographystyle{ws-procs975x65}

\end{document}